\documentclass[useAMS]{mn2e}
\usepackage{psfig}
\usepackage{graphicx}

\def\lesssim{\mathrel{\hbox{\rlap{\hbox{\lower4pt\hbox{$\sim$}}}\hbox{$<$}}}}
\def\gtrsim{\mathrel{\hbox{\rlap{\hbox{\lower4pt\hbox{$\sim$}}}\hbox{$>$}}}}

\title{Bulk Motion of Ultrarelativistic Conical Blazar Jets}

\author[Gopal-Krishna et al. ]
{Gopal-Krishna$^{1}$, Paul J.\ Wiita$^{2}$, Samir Dhurde$^{3}$ \\
$^1$National Centre for Radio Astrophysics, TIFR, Pune
                University Campus,
Post Bag No.\ 3,   Pune 411007, India; \\ \ \ \ Email: krishna@ncra.tifr.res.in\\ %}
$^2$Department of Physics \& Astronomy, P.O.\ Box 4106,
Georgia State University, Atlanta, GA 30302-4106, USA;\\ \ \ \ Email: wiita@chara.gsu.edu\\
$^3$Inter University Centre for Astronomy \& Astrophysics, Pune University
Campus, Post Bag No.\ 4, Pune 411007, India; \\ \ \ \ Email: samir@iucaa.ernet.in}
                                                                               
\date{Accepted xxx Apr 2006.
      Received xxx Mar 2006;
      in original form xxx Aug 2005}
                                                                                               
\pagerange{\pageref{firstpage}--\pageref{lastpage}}
\pubyear{xxx}
\begin{document}
\maketitle
\label{firstpage}
                                                                                               
\begin{abstract}
Allowing for the conical shape of ultrarelativistic blazar
jets with opening angles of a few 
degrees on parsec-scales
 we show that their bulk Lorentz factors and viewing angles can be
much larger than the values usually inferred by combining
their flux variability and proper motion measurements.  
This is in accord with our earlier finding that such 
ultrarelativistic (Lorentz factor, $\Gamma > 30$) conical jets 
can reconcile the relatively 
slow apparent motions of
VLBI knots in TeV blazars with the extremely fast flows implied 
by their rapid $\gamma$-ray variability.  This jet geometry also implies
that de-projected jet opening angles will typically
be significantly underestimated from VLBI measurements.
In addition, de-projected jet lengths will be considerably overestimated
if high Lorentz factors and significant opening angles are not taken
into account.
\end{abstract}
                                                                                

\section{Introduction}

Although there is a consensus about the synchrotron origin of 
the radio emission from blazar jets, the values of the bulk Lorentz factor,
$\Gamma$, and the misalignment angle from the line-of-sight, 
$\theta$, remain two key unknowns for any particular jet. Very Long Baseline 
Interferometry (VLBI) measurements of 
apparent motion of the parsec-scale radio knots have often been
employed to constrain a combination of $\Gamma$ and $\theta$ 
(e.g., Vermeulen \& Cohen 1994; Jorstad et al.\ 2001a).
The degeneracy is broken by combining these data 
with additional observations, such as flux variability, or 
high-energy photons arising from (presumably) the synchrotron 
self-Compton (SSC) mechanism, when such data are available.  
Failing this availability, the jet parameters are frequently 
estimated by simply 
setting $\theta$ equal to its most probable single value 
(i.e., $\theta = 1/ \Gamma$) (e.g., Vermeulen \& Cohen 1994;
Chiaberge et al.\ 2000). 

In all these studies, it has been customary to assume
a single value for the dominant emitting region, and we
shall also do so here, for illustrative purposes.
We note that many of the complications added
by the possibility of a range in $\Gamma$s in a given
jet have been discussed (e.g., Lind \& Blandford 1985;
Vermeulen \& Cohen 1994).  It has also been customary
to assume (often implicitly) a narrow
cylindrical geometry for the jet. As recently emphasized by us,
this is a valid assumption only provided the opening angle of 
the relativistic jet, $\omega$ is much smaller than its beaming 
angle, $\psi \sim \Gamma^{-1}$, (Gopal-Krishna, Dhurde \& Wiita 2004, hereafter 
referred to as GDW). 

It is important to note that on the sub-parsec scale, jets are
probably  
still in the collimation regime, so $\omega$ is likely to be several 
degrees, as found, e.g., for the M87 jet (Junor, Biretta \& Livio 1999).
The evidence for this conical jet hypothesis continues to grow.
Recently, Tavecchio et al.\ (2004) have provided evidence that
the only two blazar jets for which their study could be carried out
remain conical from sub-pc to $\sim$100 kpc scales.  Jorstad
et al.\ (2005) have been able to measure projected opening angles
for 15 blazar, quasar and radio galaxy jets using multi-epoch 
 VLBA images taken
at 43 GHz.  They find projected half-opening angles between
2.4$^{\circ}$ and 37$^{\circ}$ and then estimate true
full-opening angles, $\omega$, between 0.2$^{\circ}$ and 7.6$^{\circ}$,
with a mean of about 2$^{\circ}$.  

As a result of this finite opening angle, for
 small inclination angles, the Doppler boosting 
of an ultra-relativistic jet ($\Gamma \gg 10$) as well as the apparent 
proper motions, can greatly vary across the jet's 
cross-section, even when the relevant Lorentz factor for
each knot is constant. It can then be
important to carry out an integration of various quantities across 
the jet cross section.   We recently 
showed that for an ultra-relativistic jet, such a refinement can 
often result in a drastic reduction of the apparent superluminal 
motion, compared to the canonical estimates (GDW). 

We thus argued 
that it is possible to reconcile even very large values of  $\Gamma$ 
(approaching 100), which are favored by many models of
TeV $\gamma$-ray emission (e.g., Mastichiadis \& Kirk 1997;
Krawczynski et al.\ 2001) with the rarely observed presence
of apparently superluminal
motions in TeV blazar jets (e.g., Piner \& Edwards
2004; Giroletti et al.\ 2004), without invoking very 
large velocity gradients across the jet, very rapid deceleration, 
or extremely unlikely tiny viewing angles  (GDW). 
While the first of those alternatives, particularly the idea
of a spine-sheath geometry on  parsec scales (e.g., Chiaberge et al.\ 2000;
Ghisellini et al.\ 2004) retains some  attractions, the result that
blazar jets appear to retain roughly the same Lorentz factor
all the way out to multi-kpc scales when reasonable
estimates can be made (Tavecchio et al.\ 2004; Jorstad \& Marscher
2004)
militates against the generality of the deceleration picture of Georganopoulos \&
Kazanas (2003) and the extremely small viewing angle possibility is
statistically unlikely (Piner \& Edwards 2004).  
The production of ultrarelativistic velocities through a ``hydrodynamical booster'',
which is only possible in relativistic flows
has recently been shown to be viable by Aloy \& Rezzolla (2006).

In this communication, 
we present a quantitative application of our approach to the 
procedure that is usually followed to infer $\Gamma$, by combining
the apparent (often superluminal) speed, $v_{app}$, of VLBI
components with the estimated value 
of the bulk Doppler factor, $\delta$, of the jet whose axis makes an 
angle $\theta$ from the line-of-sight.  We find that the 
inferred values of both $\Gamma$ and $\theta$ are often
substantially smaller than their actual values.
We also show that the de-projected opening angle of the parsec-scale
ultrarelativistic jet frequently is considerably smaller than the physical
opening angle, and thus in accord with recent estimates (e.g.\ 
Jorstad et al.\ 2005).

\section{Conical Jet Model}

The standard expression for the Doppler factor is 
\begin{equation}
\delta = [\Gamma (1 - \beta~{\rm cos}\theta)]^{-1},
\end{equation}  %                             Eq (1)
with $v = \beta c$ and $\Gamma = (1 - \beta^2)^{-1/2}$.
The corresponding expression for the apparent speed is
\begin{equation}
\beta_{app} = \frac{\beta~{\rm sin}\theta}{1 - \beta~{\rm cos}\theta}. %Eq (2)
\end{equation}  

There are several ways of estimating $\delta$ for a blob
emitting incoherent synchrotron radiation. For instance,
when the angular size of the blob is known from VLBI, one 
can compute (i) the  {\it inverse Compton Doppler factor}, 
$\delta_{IC}$, using x-ray emission assumed to be of SSC
origin (e.g., Marscher 1987; Ghisellini et al.\ 1993), 
or (ii) the {\it equipartition Doppler factor}, $\delta_{eq}$, 
using the radio spectral turnover (Readhead 1994; Guerra \& Daly 1997; 
also, Singal \& Gopal-Krishna 1985).  If VLBI data at
multiple epochs for multiple components is available, 
a $\delta_{\beta_{app}}$ can be
found by assuming that the highest $\beta_{app}$ defines the
minimum Lorentz factor, so $\delta_{\beta_{app}} \sim \beta^{max}_{app}$
(e.g., Jorstad et al.\ 2005).  

Alternatively, $\delta$ can be estimated from radio 
observations of flux variability associated with a new VLBI 
component (``knot''), by adopting some maximum physically
attainable value for 
the intrinsic brightness 
temperature, $T_{max}$ (e.g., Valtaoja et al.\ 1999). This $T_{max}$ 
 could be set either by the equipartition condition 
($\sim 5 \times 10^{10}$ K, Readhead 1994; Singal \& Gopal-Krishna 1985), 
or by the inverse Compton 
catastrophe ($\sim 10^{11} - 10^{12}$ K, Kellermann \& Pauliny-Toth 
1969). 
If an appropriate variability timescale, $\tau_{obs}$, is found
corresponding to an observed flux variation $\Delta S$ measured
at a frequency $\nu$, then (ignoring cosmological effects)
$T_{B,obs} \propto \Delta S/(\tau_{obs} \nu)^2$ and 
$\delta = (T_{B,obs}/T_{max})^{1/(3+\alpha)}$,
where $\alpha$ is the spectral index ($S_{\nu} \propto \nu^{-\alpha}$) 
(e.g., Ter\"asranta \&
Valtaoja 1994).
This last method, which actually produces a lower bound to
$\delta$, or  $\delta_{min}$, has 
been used quite commonly because it does not require
VLBI measurements (e.g., Fanti et al.\ 1983; Singal \& Gopal-Krishna 1985;
Ter\"asranta \& Valtaoja 1994).

The foregoing equations (1) and (2) can be combined to solve for $\Gamma$ and 
$\theta$ of the knot, assuming a cylindrical jet (e.g., Guerra \& Daly 1997), 
in terms of $\beta_{app}$ and $\delta_{min}$,
\begin{equation}
\Gamma =  \frac{\beta_{app}^2 + \delta_{min}^2 +1}{2 \delta_{min}}, 
\end{equation}    %       Eq (3)
and
\begin{equation}
{\rm tan}~ \theta = \frac{2 \beta_{app}}{\beta_{app}^2 + \delta_{min}^2 -1}. 
\end{equation}  %           Eq (4)
Within the context of this last method for estimating $\delta_{min}$,
we shall now proceed to quantify how the solutions for $\Gamma$ and
$\theta $ (Eqs.\ 3 and 4) are affected when an allowance is 
made for the jet's conical geometry (with a finite full opening angle, 
$\omega$), which, as mentioned 
above, can be several degrees on parsec scales.  

The procedure for computing the mean flux boosting factor, 
$\bar{A}$, which is the same as the brightness temperature
boosting factor, such that the
observed flux $S_{o,w} = \bar{A} S_{em}$, with $S_{em}$ the
emitted flux, has been described in GDW.  Also described there
is the way in which the boosting weighted effective apparent 
speed, $\beta_{app,w}$, is found by integrating over the jet cross section.  
 The effective value
(that which would be observed)
of $\delta_e$  is
\begin{equation}
\delta_{e} = (\bar{A})^{1/(3+\alpha)},
\end{equation}
where we have assumed that the emission is concentrated in the knot;
if the emission were from a continuous flow the integer in the
exponent in Eq.\ (5) would be 2 instead of 3.
In the following, we have made the common assumption of a flat radio spectrum 
for the VLBI knots ($\alpha = 0$), though our main conclusions are not sensitive
to reasonable deviations from  this flatness assumption for core
dominated sources.

These effective parameters can now be used to compute the 
values of $\Gamma_{\inf}$ and $\theta_{\inf}$ that would be inferred
from the standard formulae (3) and (4).  These inferred values
can then be compared with the actual intrinsic values
of $\Gamma$ and $\theta$ adopted for the jet while
  computing $\delta_e$ and $\beta_{app}$.

\section{Results}

Taking several combinations of $\Gamma$ and $\omega$ for jets, 
we plot in Fig.\ 1 the computed $\theta$-dependences
of the values of $\Gamma_{\inf}$ that would be {\it inferred} 
from flux variability and VLBI proper motion
data in the conventional approach where 
the angular width of the (parsec-scale) jet is ignored 
by assuming a cylindrical geometry. Alongside these $\Gamma_{\inf}$ 
curves, 
we also show the probabilities,
$p(\theta)$, of observing a jet in a flux limited sample
at a viewing angle of $\theta$. All curves are computed using
a grid spacing of 0.01$^{\circ}$ within the jet's crosssection.
%we plot the probability
%weighted values of $\Gamma^{\prime}_{inf}$. 
%We also show the expectation
%value of $\Gamma^{\prime}_{inf}$.
%by weighting them, at different
%$\theta$, with $p(\theta)$.

\begin{figure*}
\hspace*{-0.25cm}\psfig{file=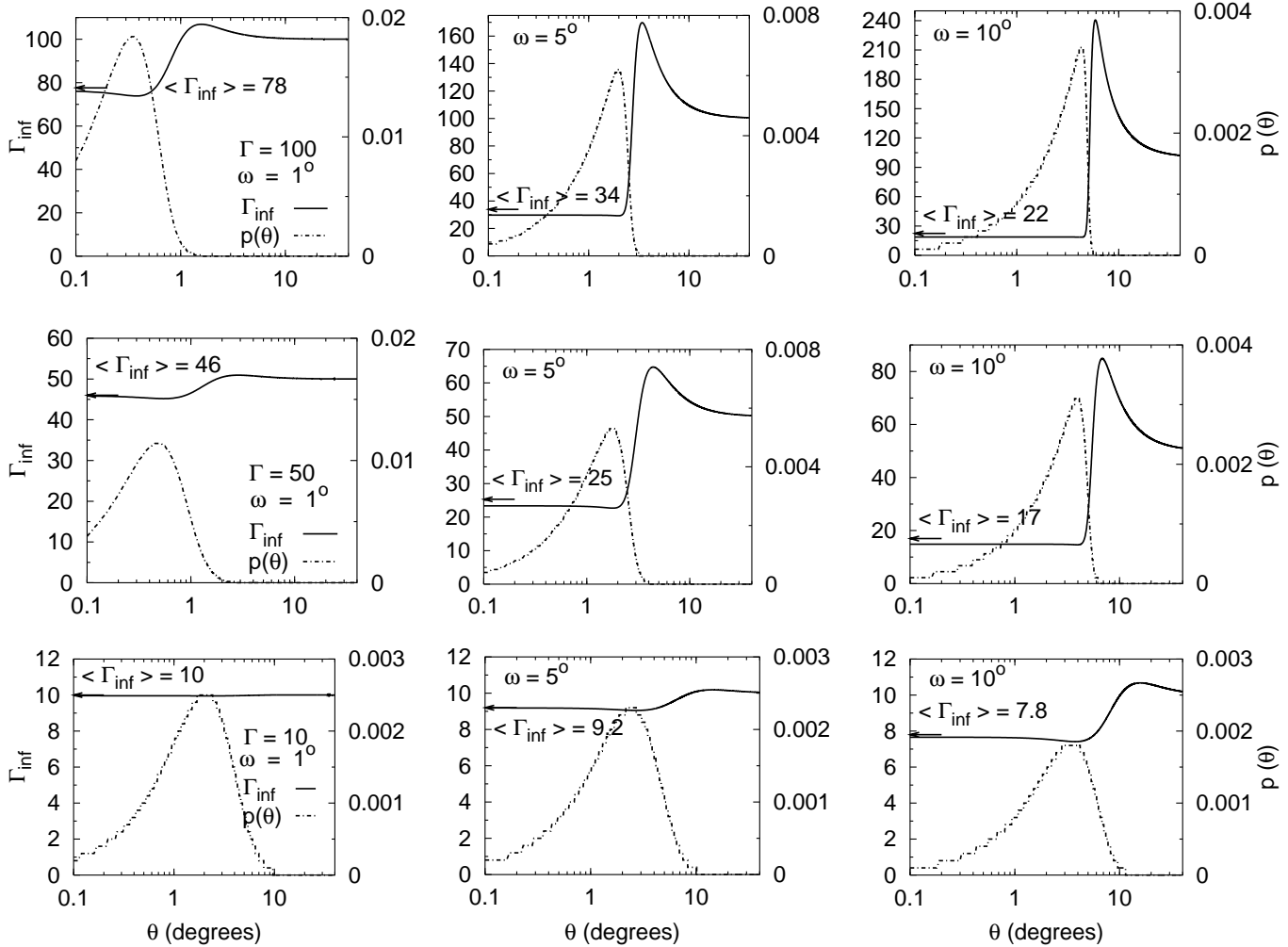}
\caption{Inferred Lorentz factors, $\Gamma_{\inf}$, 
(solid curves) and flux-limited sample 
probability functions, $p(\theta)$,  
(dot-dashed curves) against the viewing 
angle to the jet axis, $\theta$.  The top row has
$\Gamma = 100$, the middle row, $\Gamma = 50$ and
the bottom row, $\Gamma = 10$.  Each panel
is labelled with the actual value of the full opening
angle $\omega$.  Also noted with arrows are the expectation values,
$\langle \Gamma_{\inf} \rangle$.}
\end{figure*}

Following GDW, this probability function, $p(\theta)$, 
is given by 
\begin{equation}
p(\theta)d\theta \propto {\rm sin}\theta A^q(\theta) d\theta,                
\end{equation}         %            Eq (5)
where $q \simeq 3/2$ is the slope of the integral source counts 
(Log $N$ -- Log $S$) at centimetre wavelengths (e.g., Kapahi 1987;
Cohen 1989).
We also show in Fig.\ 1 the expectation
value, $\langle \Gamma_{\inf} \rangle$,  which is found by weighting the values
of $\Gamma_{\inf}(\theta)$  by $p(\theta)$.

In Fig.\ 2 we plot  values of $\theta_{\inf}$ against $\theta$,
along with the
expectation values of $\langle \theta_{\inf} \rangle $, which are also computed
for a flux-limited sample using the same $p(\theta)$ distributions.
As expected, for small enough $\omega$,  about $1^{\circ}$,
the inferred values of $\Gamma$ and $\theta$ are both quite close
to the adopted intrinsic values, although 
even then for very high $\Gamma$ a difference at the $\sim$20 per cent 
level for $\Gamma_{\inf}$
is seen.  

\begin{figure*}
%\hspace*{-0.2cm}
\psfig{file=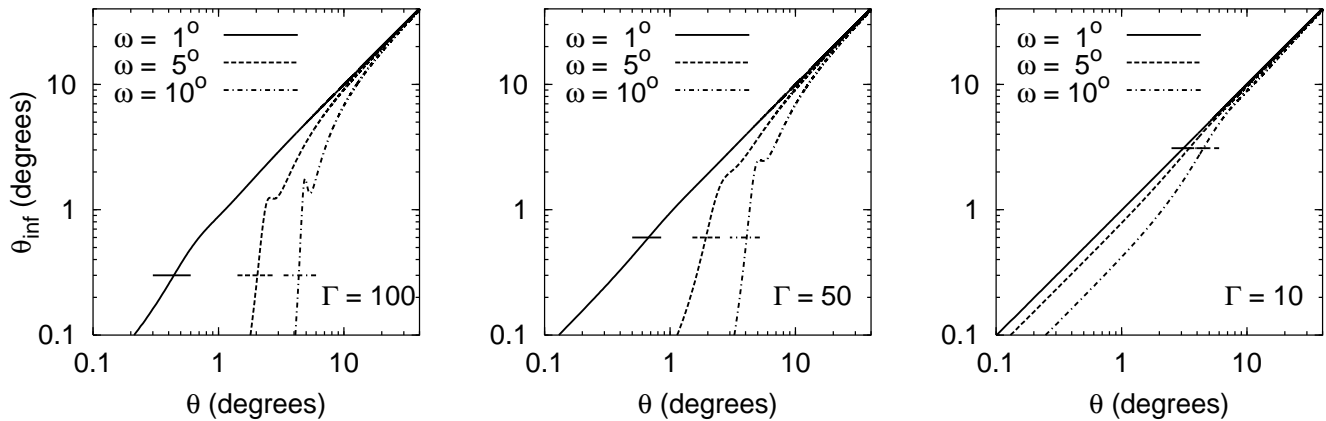}
\caption{Inferred viewing angles, $\theta_{\inf}$,
 against  $\theta$.  Each panel
is labelled with the actual values of $\Gamma$ with curves 
displayed for
 three values of the full opening angle  $\omega:  1^{\circ}$ (solid),
$ 5^{\circ}$ (dashed)
and $10^{\circ}$ (dot-dashed). The expectation values,
$\langle \theta_{\inf} \rangle$ for those 
values of $\omega$ are shown by similar types
of  line-segments.}
\end{figure*}

The characteristic behaviour of $\Gamma_{\inf}$ can be segmented
into three regimes.  For $\theta$ less than some critical angle,
$\theta_c \simeq \omega/2$, $\Gamma_{\inf}$ remains essentially
constant at a value which can be much smaller than $\Gamma$.
The computed expectation value is dominated by this reduced
$\Gamma_{\inf}$, since for $\theta > \theta_c$ the probability
of viewing such a source, $p(\theta)$, drops drastically (Fig.\ 1).  
Approaching the critical viewing angle from below,
a sharp rise in $\Gamma_{\inf}$ to a value exceeding $\Gamma$
is found, the excess being more pronounced
for larger $\omega$.  Finally, at still larger $\theta > \theta_c$,
$\Gamma_{\inf}$ declines and asymptotically approaches $\Gamma$;
however, the chance of seeing a source in either of these last
two regimes is very small.
This behaviour is a basic consequence of the spatial sharpness of
the region of strongest Doppler boosting, across which the 
gradients of $\beta_{app}$ and $A$ can be positively or negatively
correlated.  Another key factor is how much of this region 
(whose size is $\sim 1/\Gamma$) is encompassed within the jet's
crosssection at a given viewing angle.  Clearly, for large $\Gamma$,
this coverage is very  sensitively dependent on $\theta$ when 
$\theta$ approaches $\theta_c$.  In particular, as the periphery
of the jet's crosssection crosses over the line-of-sight ($\theta < \omega/2$)
the value of $\beta_{app}$ averaged over the cross-section
will decrease from a value which was comparable to $\Gamma$
(and will become 0 when $\theta = 0$), leading to a fall in 
  $\Gamma_{\inf}$ (Eq.\ 3). 
%At the same time, however,
%there will be a steadily increasing contribution to the Doppler boosted flux,
%and hence to $\delta_e$, so that $\Gamma_{\inf}$ will be reduced
%well below $\Gamma$ (Eqs.\ 1, 3).

In Fig.\ 2 the most noticeable feature is the sharp drop in
$\theta_{\inf}$ as as $\theta$ approaches $\theta_c$ from
above.  This can also be understood because at this viewing
angle the jet's crosssection begins to encompass the narrow region
close to the line-of-sight ($\theta < 1/\Gamma$) where the
Doppler boosting is most extreme.  Thus the flux-weighted
viewing angle of the beam, $\theta_{\inf}$, drops to very
small values.  
For $\theta > \theta_c$ the
values of $\theta_{\inf}$ begin to approach $\theta$ itself, and
once $\theta > \omega$ the inferred viewing angle barely
differs from the actual one.
The expectation values, $\langle \theta_{\inf} \rangle$, are 
found to be nearly independent of $\omega$, and scale 
approximately (in radians) as $0.5/\Gamma$, somewhat less than the 
most probable value for $\omega = 0$, $1/\Gamma$.
To summarize, the standard procedure of estimating $\Gamma$ and
$\theta$ from $\delta$ values obtained from flux variability measurements can 
grossly underestimate their values if the jets
are highly relativistic and have modest opening angles.
Often the standard procedure may yield implausibly
precise alignment ($\theta_{\inf} \ll 1^{\circ}$) even
when the true viewing angle (to the axis of the jet)
is a few degrees.

Another important parameter that may also be substantially underestimated
is the width of the VLBI knots, thereby making the pc-scale jet appear
much better collimated than it really is.  For instance,
Jorstad et al.\ (2005) find that most of their 7mm VLBA observations
indicate full de-projected opening angles of only 1--4$^{\circ}$.   However,
these could be underestimated, since for an 
ultrarelativistic jet the Doppler boosting is extremely
pronounced for the portion of the jet's cross-section which
lies within an angle of $\sim 1/\Gamma$ to the line-of-sight.  
To demonstrate this
quantitatively, in we plot in Fig.\ 3 the FWHM, FWQM, and full-width
at one-tenth maximum (FW(0.1)M)
of the Doppler boosting
distribution, thereby providing estimates for the  de-projected opening
angles, $\omega_{\inf}$, that are expected to be
inferred from VLBI data.  These are computed in an 
approximation only valid for ultrarelativistic jets when
$\omega > 1/\Gamma$.
%The expectation values, $\langle \omega_inf \rangle$ 
%(for the FWQM) are roughly equal to $2/\Gamma$
%for these cases.
We have plotted these results only for $\omega = 5^{\circ}$ since the
corresponding plots for $\omega = 10^{\circ}$ are nearly the same.
As expected,  these inferred opening angles basically scale with $\Gamma^{-1}$.
Note that these computed $\omega_{\inf}$ values are slight underestimates,
as they should be convolved with the
beaming pattern of each element of the jet's cross-section, which would
widen the FWHM to a minimum value of $2/\Gamma$.
%Also  across jets as functions of $\theta$ for various 
%values of $\Gamma$ and $\omega$, as well as the expectation
%values of the FWHMs.  When $\theta > 1/\Gamma$ then the inferred
%angular width, $\omega_{inf}$ is found to saturate near 
%$1/\Gamma$ (about 1$^{\circ}$ for
%$\Gamma = 50$), even if the actual $\omega$ is substantially larger.
%CHECK THIS AGAINST NEW PLOTS

\begin{figure}
\hspace*{0.4cm}\psfig{file=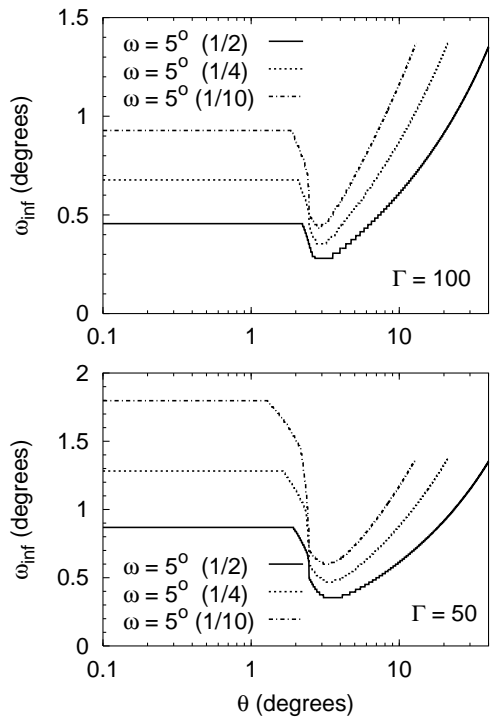}
\caption{Inferred full opening angles, $\omega_{\inf}$,
shown as FWHM (solid), FWQM (dotted), and FW(0.1)M (dot-dashed)
estimates  against  $\theta$,
for $\Gamma = 100, 50$ and $\omega =5^{\circ}$.}
\end{figure}

\section {Discussion and Conclusions}

In our earlier paper (GDW) we quantitatively checked the consistency 
with the VLBI proper motion data of the hypothesis that, on the 
parsec-scale, many blazar jets are extremely relativistic ($\Gamma > 30$) 
and have full opening angles of several degrees.  In this paper we
argue for the consistency of such high Lorentz factors with
observations of flux variability, by showing that standard
analysis of such data often 
leads to underestimation of  $\Gamma$ and $\theta$ if the conical 
shape of the jets is not properly taken into account. Ultrarelativistic jets
have been independently inferred recently from $\gamma$-ray variability 
(e.g.\ Ghisellini et al.\ 2005; Krawczynski, Coppi \& Aharonian 2004) 
and modelling of TeV blazar spectra (e.g.\ Krawczynski 2004, and references therein)
as well as rapid radio variations (Begelman, Ergun \& Rees 2005, and 
references therein) and very high $T_B > 10^{13}$ K deduced from space 
VLBI (e.g., Horiuchi et al.\ 2004; Kovalev et al.\ 2005).  In our earlier work we showed that
the frequently observed subluminal or mildly superluminal apparent 
motions of TeV blazar VLBI knots can be reconciled with 
ultrarelativistic jets provided they have a roughly conical
geometry with modest full opening angles, $\omega$, of 5--10$^{\circ}$
on the parsec scale.

Here we show that such characterizations of jets can also
explain the smaller values of $\Gamma$ and $\theta$ that 
are usually inferred by combining the observations of apparent 
motions of VLBI knots with their radio flux variability under 
the assumption of a cylindrical jet ($\omega \rightarrow 0^{\circ})$.
We note from Fig.\ 1 that the expectation values of the inferred $\Gamma$
for $\Gamma = 50$ to $100$ and $\omega = 5^{\circ}$ to $10^{\circ}$, are in the range 
17--34.  While these values for $\Gamma_{\inf}$ are still roughly
twice those typically inferred from the analysis
of variability data of blazars (e.g. Ter\"asranta \& Valtaoja 1994),
since the latter estimates of Lorentz factor
are only lower limits, via a $\delta_{min}$, there need
not be any contradiction.  Our values for $\Gamma_{\inf}$
are not much above those found by Jorstad et al.\ (2005)
who use a modified variability technique which compares the
timescale of decline of flux density with the light travel
time across the emitting region.
Likewise, we have shown that the tendency of VLBI jets to appear quite well
collimated (with $\omega_{\inf} \sim 2^{\circ}$, Jorstad et al.\ 2005) 
can also be 
reconciled with our assumption of conical jets with opening angles which are
actually a few times larger.

For TeV blazars, two key arguments made for strong deceleration of jets
between the TeV emitting sub-parsec scale to the radio emitting parsec scale are:
(i) The frequently observed mildly relativistic or sub-relativistic motion
of the VLBI knots (e.g., Piner \& Edwards 2004);
(ii) the problem with
FR I--BL Lac unification, since  $\Gamma \ga 10$ would grossly overpredict the
number of FR I galaxies above a given limiting flux  (e.g., Hardcastle et al.
2003).  We stress that the conical jet geometry could resolve both of
these problems for non-decelerated flows (see below).

Piner \& Edwards (2004) have reported that the few known TeV blazars display 
at most mildly superluminal VLBI knots, in statistically significant contrast 
to normal blazars, which show both mildly and truly superluminal radio
knots (Jorstad et al.\ 2001a,b). To address this 
intriguing absence of relativistic shocks in TeV blazars, Piner \& Edwards
considered the alternative possibility that the apparent subluminal motion 
in TeV blazars could be due to a closer alignment of the jet to the line-of-
sight. However, since the required angle is $\lesssim 1^\circ$, they discount this 
scenario on statistical grounds and  instead favoured the possibility that 
the jets actually decelerate to modest bulk Lorentz factors between sub-parsec
and parsec scales. (However, this raises the question: what renders the
deceleration mechanism less effective in the case of normal blazars?)
In GDW we have shown that the slow apparent motion of 
VLBI knots can be an artefact resulting from the finite opening angle of the 
jet.  From 
our present analysis (Sect.\ 3) it is further evident that the jet misalignment angle
required to account for mildly relativistic or non-relativistic apparent motion
can be as large as a few degrees (instead of $<$1 degree). This is statistically
much more plausible, so the hypothesis of the jet's rapid deceleration is not
necessary on this account.

The second difficulty, related to the parent population of BL Lacs,
can be addressed by recalling that the beaming 
angle for an ultra-relativistic conical jet would at least equal its opening 
angle (even for an idealized case when the bulk motion of the radiating 
plasmons within the jet is ballistic, i.e., purely along straight 
trajectories). Thus, the effective beaming angle of an ultrarelativistic jet 
being considered here ($\Gamma \sim 50$, $\psi \sim 10^\circ$), could easily 
correspond to that expected for a $\Gamma \sim 5$ cylindrical jet and hence 
be consistent with the FR I--BL Lac unification scheme (e.g., Urry \& Padovani 
1991; Hardcastle et al.  2003). For misalignment angles larger
than the cone angle, it is conceivable that appreciable flux is received from
a core component of the type detected in the nearby radio galaxy M87, which
is non-varying (Kovalev et al.\ 2005) and hence probably at most mildly beamed. 

In this context, it may  be worthwhile to recall the cases of two blazars 
where modelling of the broadband spectra has yielded no evidence for 
deceleration of highly relativistic jets from subparsec scales to distances 
of hundreds of kiloparsecs (Tavecchio et al., 2004).

We note that our aim in the present study is not to reject the
possibility of a drastic deceleration taking place between sub-parsec and 
parsec scales, accompanied with the formation of spine-sheath velocity 
structure in parsec scale jets, as inferred, e.g., by Chiaberge et al.\ (2000) 
(also, Marscher 1999; Trussoni et al.\ 2003).   We merely seek to 
question the generality of this scenario as well as the underlying tacit 
assumption that the process responsible for the creation of radio emitting
knots (e.g., some instability whose source lies at the base of the jet) operates
almost exclusively within the (slower) sheath region. We have argued that even when
radio knots form within the putative ultrarelativistic spine, this can 
be in accord with the observed subluminal motions without resorting to the 
assumption of a very close jet alignment, which is highly implausible. 
Still, one expects the jets in TeV blazars to be fairly well directed
towards us, owing to the dependence of the $\gamma-$ray flux on a high
power of $\delta$ (cf.\ Piner \& Edwards 2004).
Our model does predict that a small fraction of TeV blazars
should show strongly superluminal motions (GDW).  However, in light of the modest number of
such sources known so far, and the smaller number which have had their
structure mapped frequently with VLBI, the fact that no such apparent velocities have been found
is to be expected.  As more such blazars are discovered by HESS, MAGIC and
VERITAS, the sample on which the VLBI studies can be performed should become
large enough to test this prediction.
                                                                                  
The situation appears to be much less intriguing for normal blazars, as seen 
from two large samples observed at cm/mm wavelengths using the VLBA (Jorstad 
et al.\ 2005; Kovalev et al.\ 2005). While a sizable fraction of these blazars
does exhibit subluminal motion, Lorentz factors estimated for the vast majority
fall within the range 5 to 40 (e.g., Jorstad et al.\ 2005). Similarly large
values of $\Gamma \ga 30$ have been estimated in the past from VLBI monitoring
of a few well known blazars (e.g., Fujisawa et al.\ 1999). Note that the 
occurrence of subluminal radio
knots in a small fraction of normal blazars is fully consistent with the present
model invoking modestly misaligned ultrarelativistic conical jets (GDW).
                                                                                     
We note that recent VLBI studies have also revealed several cases
of ultra-bright radio knots, with T$_B$ extending up to (at least) $5 \times 10^{13}$ K (Kovalev et
al.\ 2005; Horiuchi et al.\ 2004). In addition, fairly robust estimates of
angular sizes have been made for a few blazars from their intra-day flux
variations at centimetre wavelengths, yielding brightness temperatures of
$5 \times 10^{13}$ K or more (Rickett et al.\
2002; Macquart \& de Bruyn 2006), which cannot be understood in terms of a combination of
modest Lorentz factors and refractive interstellar scintillation. 
As discussed by these authors, interpretation of at least
these ultra-bright sources in terms of steady incoherent synchrotron radiation
would still demand extremely large bulk Lorentz factors ($\Gamma \ga 100$).
The only way to avoid such large $\Gamma$ values appears to be to invoke some coherent
radiation mechanism (e.g., Begelman et al.\ 2005).

One disadvantage of the ultrarelativistic scenario is a decreased efficiency in the 
conversion of bulk energy to radiated energy 
(e.g. Begelman, Rees \& Sikora 1994); however, this problem is, on average, no more 
severe for this conical jet picture than it is for a cylindrical jet with the same $\Gamma$.  
If one compares two jet knots with identical intrinsic luminosities and emissivities,  both with $\Gamma =50$,
the ratio of the observed flux from a jet with full opening angle $\omega = 5^\circ$ to that 
observed 
from the cylindrical equivalent clearly
depends on the (actual) viewing angle, but the range is not very dramatic.
This ratio, $\bar{A}/\delta^3$, is: 0.17 at $\theta = 0.5^\circ$; 0.54 at $\theta = 1.0^\circ$;
5.09 at $\theta = 2.0^\circ$;  2.93 at $\theta = 5.0^\circ$; and is 1.0 for $\theta \ge 15^\circ$.
We have already shown (GDW) that the limit on total jet power set by a comparison between the 
Eddington luminosity and the inferred bolometric luminosity is not a significant constraint.                        

The presence of ultrarelativistic bulk motion in the VLBI jets, 
as argued here to be in accord with a variety of observations, 
have other interesting observational implications.
For instance, the deprojected length of jets as well as the
radio lobe separation could be substantially overestimated, since the actual
viewing angle is often much larger than the $\theta_{\inf}$ inferred by
assuming the jet to be cylindrical (Sect.\ 3).   The substantial reduction in the deprojected lengths of the jets, as
argued here, has important ramifications in several contexts, one of which
is the origin of optical and x-ray emission associated with radio hot spots.
(e.g., Gopal-Krishna et al.\ 2001; Prieto et al.\ 2002; Brunetti et al.\ 2003).
Further, the possibility of substantial underestimation of viewing angle
can ease the discomforting inference that exceptionally large radio lobes 
are associated with several prominent blazars (Schilizzi \& de Bruyn 1983).

In addition, the inverse Compton boosting of the UV photons from
the accretion disk by $\Gamma \sim 10$ jets would lead to an 
additional contribution to the SED in the soft x-ray band, 
the so-called ``Sikora bump''; its absence in blazar spectra 
has been used to argue against lepton dominated relativistic jets on the 
parsec-scale (Sikora \& Madjeski 2000). On the other hand, for 
$\Gamma \sim 50-100$ jets the corresponding bump would be
pushed up to very hard x-rays where it is presently much more difficult to detect,
and so the lack of the soft x-ray Sikora bump does not currently constrain such models.  
We finally note that ultrarelativistic jets containing shocks
with finite opening angles, similar to those discussed
here for blazars, are usually invoked in modelling the afterglows 
from Gamma Ray Bursts (e.g., Sari, Piran \& Halpern 1999;
Panaitescu \& Kumar 2000; Frail et al.\ 2001).
Thus the plausibility of very high Lorentz factors of blazar
jets hints at an underlying similarity between the jets of
GRBs and blazars.  

\section*{Acknowledgments} 
We thank the anonymous referee for suggestions which significantly
improved the discussion of our results.
SD  is grateful to NCRA for the use of its research facilities. 
PJW appreciates the hospitality provided at both NCRA and 
the Department of Astrophysical Sciences, Princeton
University.  PJW's efforts were partially supported by a subcontract to Georgia
State University from US NSF 
grant AST-0507529 to the University of Washington.

\end{document}